\documentclass{elsart}
\usepackage{epsfig,latexsym,amsmath}
\usepackage[figuresright]{rotating}
\def\bear{\begin{eqnarray}}
\def\eear{\end{eqnarray}}

\def\be{\begin{equation}}
\def\ee{\end{equation}}
\def\bear{\begin{eqnarray}}
\def\eear{\end{eqnarray}}
\def\bse{\begin{subequations}}
\def\ese{\end{subequations}}

\begin{document}
\begin{frontmatter}

\title{Estimation of System Parameters and Predicting the Flow Function 
 from Time Series of Continuous Dynamical Systems}

\author{P.~Palaniyandi\thanksref{mail1}} and
\author{M.~Lakshmanan\thanksref{mail2}}

\address{Centre for Nonlinear Dynamics, Department of Physics
Bharathidasan University, Tiruchirapalli 620 024, India}
\thanks[mail1]{palani@cnld.bdu.ac.in}
\thanks[mail2]{lakshman@cnld.bdu.ac.in}
\date{}

\begin{abstract}

We introduce a simple method to estimate the system parameters in continuous
dynamical systems from the time series. In this method, we  construct a
modified system by introducing some constants (controlling constants) into the
given (original) system. Then the system parameters and the controlling
constants are determined by solving a set of  nonlinear simultaneous algebraic
equations obtained from the relation connecting original and modified systems.
Finally, the method  is extended to find the form of the  evolution equation of
the system itself. The major advantage of the method is that it needs only a
minimal number of time series data  and is applicable to dynamical systems of
any dimension.  The method also  works extremely well even in the presence of
noise in the time series.  This method  is illustrated for the case of  Lorenz
system.

\end{abstract}


\end{frontmatter}

Time series analysis (both vector and scalar) is of considerable relevance
\cite{abar,kantz,diks} to physical, chemical and biological systems as they
very often exhibit temporal chaotic motions. The main objective of time series
analysis is to study the detailed structure of the evolution equation of the
underlying dynamical system.  This includes the number of independent
variables, the form of the flow functions and parameters involved in the
system.  In this Letter, the study is focussed on the last aspect, namely, 
estimating the system parameters from the  time series when partial
information  about the system is known (the number of independent variables and
the form of the flow functions), and then the study is extended to predict the
form of the flow function itself when it is not known.   A few methods have
been proposed \cite{abar,kantz,diks} in the literature for predicting the form
of the flow functions.  These include two point boundary value problem approach
\cite{e1}, Euler integration approach to odes \cite{abar} and modifications
\cite{e2}.  Also recent literature shows that the methods already proposed for
estimating the system parameters are based on the concept of synchronization
\cite{p1,p2,p3,p4,p5}, Bayesian approach \cite{p6,p7},  least squares approach
\cite{p8} and by successive integration method \cite{p9}. In this Letter, a
very simple and efficient method for estimating the system parameters as well
as the form of the flow functions of continuous dynamical systems from the 
vector time series is developed using  the concept of controlling chaos
\cite{mlsrbook,mlkm,sp}, which can also in principle be extended to scalar time
series. In our method we construct a modified system by inclusion of certain
constants (controlling constants) in the given original system so that the
evolution of the modified system is controlled to an equilibrium point. Then we
find the dynamical relation between the original and  modified systems and
thereby determine the unknown system parameters and the controlling constants. 
After accomplishing this task, the method is extended to determine the form of
the evolution equations (flow functions) itself for the system from which the
time series was collected. This method is applicable to time series obtained
from a continuous system of any dimensions and is also well suited  for
discrete dynamical systems as shown in ref. \cite{pp} earlier. The method can
also be used for the time series which contains considerable amount of noise.
Further this method can be used in the field of controlling chaos to find the
exact values of controlling constants to make the given chaotically evolving
system to be controlled to a required equilibrium point.

Consider an arbitrary $N$-dimensional continuous chaotic dynamical system
(the original system),
\bear 
\dot{x}_i=f_i(x_1,x_2,...,x_N;{\bf p}),	
\label{org_sys}	
\eear 
where $i=1,2,3,...N$, and ${\bf p}$ denotes the system parameters  of dimension
$M$ to be determined.  Here we assume that the function $f$ is sufficiently
smooth. Let us construct a modified continuous dynamical system (the modified
system) as
\bear 
\dot{y}_i=f_i(y_1,y_2,...,y_N;{\bf p})+\kappa_i, \;\;   i=1,2,...,N,
\label{mod_sys}
\eear
where $\kappa_i$'s are constants (controlling constants).  The crucial idea
here is that the Jacobian matrix which determines the stability of the
equilibrium point is the same for both the cases, namely the original and the
modified systems.  In fact, the inclusion of $\kappa_i$'s in Eq.
(\ref{mod_sys}) makes the modified system to have an equilibrium point (either
stable or unstable) which is effectively different from the equilibrium point
(unstable) of the original system eventhough the Jacobian matrix is not
affected as stated above.  From an examination of many maps and flows we have
found that there is in general a possibility of making the modified system
(\ref{mod_sys}) to exhibit a stable equilibrium point by suitable inclusion of
constants  $\kappa_i$ eventhough the original system (\ref{org_sys}) evolves
chaotically.

Let $u_i(t)$ be the deviation, that is, $u_i(t) = y_i(t) - x_i(t)$, of the
trajectory of the modified system from that of the original system due to the
effect of controlling constants ($\kappa_i$'s). After substituting the above
relation in Eq. (\ref{mod_sys}) and making use of Taylor expansion, we obtain 
\bear
\dot{u}_i &=& \kappa_i + \sum^N_{j=1}u_j\frac{\partial f_i}{\partial x_j} 
\bigg\vert_{{\bf x}}   	
 +\frac{1}{2} \sum^N_{j=1}\sum^N_{k=1}u_j u_k  
\frac{\partial^2 f_i}{\partial x_j\partial x_k} 
\bigg\vert_{{\bf x}}  +\cdots, \;\;   i=1,2,...,N.
\label{dif_sys}
\eear
Here the noteworthy point is that the above equation contains  {\bf x} whose
evolution is characterized by Eq. (\ref{org_sys}).   Thus, while obtaining the
solution to Eq. (\ref{dif_sys}) one has to solve Eqs. (\ref{org_sys}) and
(\ref{dif_sys}) simultaneously.  

Let us now consider the time evolution  of the variables $x_i(t)$, $y_i(t)$ 
and $u_i(t)$, statisfying Eqs. (\ref{org_sys})-(\ref{dif_sys}), respectively,
at discrete time intervals.  For this purpose, we introduce the notation
$x^{(j)}_i = x_i(j \Delta t)$, $y^{(j)}_i = y_i(j \Delta t)$, and $u^{(j)}_i =
u_i(j \Delta t)$, where $\Delta t$ is a small time interval and 
$j=0,1,2,3,...$.  With this notation, the relation between the original and
modified systems at definite intervals can be written as  
\bear y^{(j)}_i=x^{(j)}_i+u^{(j)}_i, \;\;\;\;\; j=0,1,2,... 
\label{rel_org_mod} 
\eear

After the transient time $k \Delta t$, let the modified system approaches an equilibrium
point $y^*_i$ and hence the above equation becomes,

\bear
x^{(j)}_i+u^{(j)}_i=y^*_i, \;\;\;\;\; j=k+1,k+2,k+3,... 
\label{eqi_point_x}
\eear

Let  $ z^{(0)}_i,z^{(1)}_i,...,z^{(m-1)}_i$  be the $m$ set of data points of
the given chaotic time series sampled at the time interval $\Delta t$ from the
original system (\ref{org_sys}).  Substituting this in the above equation, we
get 
\bear 
z^{(n)}_i+u^{(n)}_i=y^*_i, \;\;\;\;\; n=0,1,2,...,(m-1).
\label{eqi_point} 
\eear 

In the above relation, it is instructive to note that we need not bother about
the transients because the time series data is collected only after sufficient
transient time.  So, the discrete time index ($n$) now can start from $0$, that
is from the first data of the time series. It may be noted that here, $u_i$'s 
are functions of the parameters ${\bf p}$ and controlling constants $\kappa_i$,
since the derivatives of $u_i$'s are having   specific functional relations
with the same parameters and controlling constants through Eq. (\ref{dif_sys}).
Also, $u^{(n)}_i$'s  are obtained by solving Eq. (\ref{dif_sys}) numerically
with time interval $\Delta t$ (or submultiples of $\Delta t$).  In general, for a given set of time series data
collected from the system (\ref{org_sys}) for a particlular set of parameters,
the right hand side of Eq. (\ref{eqi_point}), that is, the equilibrium point,
depends on the value of controlling constants ($\kappa_i$'s) which can be
varied by means of $u^{(0)}_i$.  This freedom allows us to  make any desired
point in the region of the attractor of the system as equilibrium point
($y^*_i$) by choosing the values for $u^{(0)}_i$ accordingly. For example,  one
can easily have $z^{(0)}_i$ as an equilibrium point by setting $u^{(0)}_i=0$ in
Eq. (\ref{eqi_point}). Similarly, an arbitrary point $q_i$ in the attractor can
be chosen as an equilibrium point if we start with $u^{(0)}_i=q_i-z^{(0)}_i$. 

Now it is possible to obtain the required number ($M+N$, $N$ for $\kappa_i$'s
and $M$ for ${\bf p}$'s) of functional relations (implicit) between the unknown
parameters and controlling constants through Eq. (\ref{eqi_point}). Once we
have the required number of functional relations, the next task is to solve
them for the unknowns, that is for $\kappa_i$'s and ${\bf p}$'s, using a
suitable numerical technique, for example by the globally convergent Newton's
method ~\cite{nm}.    

\begin{sidewaystable}  
\caption{Convergence of $\sigma$, $\rho$, $b$, $\kappa_1$, $\kappa_2$and $\kappa_3$ in the Lorenz system}   
\begin{center}
\begin{tabular}{|c|c|c|c|c|c|c|}
\hline 
No. & $\sigma$   & $\rho$     & $b$         &  $\kappa_1$ & $\kappa_2$  & $\kappa_3$	 \\
\hline
 0 &   1.00000000 &   1.00000000 &  1.00000000 &   1.00000000 &    1.00000000  &  1.00000000 \\
 1 &  25.06611000 & -53.43927900 &  5.90702312 &  -1.09685478 &   -1.64477291  &  5.66012411 \\
 2 &  33.88597350 &   5.87195581 &  4.74703140 &  -1.70082489 &   -2.33269811  &  7.14465601 \\
 3 & -24.31336050 &  20.11146620 &  3.81209093 &  -3.00648132 &   -3.72360435  &  9.50270473 \\
 4 &  31.73494710 &  37.21100300 &  1.76763806 &  -7.61414582 &   -9.53379493  & 19.84796380 \\
 5 & -11.56754590 &  20.78649710 &  3.33689508 & -15.06851440 &  -19.96489090  & 38.24238610 \\
 6 &   8.87479048 &  29.65454090 &  2.67975225 & -15.31294170 &  -19.87467490  & 38.27824720 \\
 7 &   9.95266093 &  27.99241480 &  2.66686356 & -15.26076520 &  -20.11294050  & 38.27375090 \\
 8 &  10.00000000 &  28.00000000 &  2.66666667 & -15.25561980 &  -20.11136290  & 38.27364540 \\   
9 & 10.00000000 &  28.00000000 &  2.66666667 & -15.25561980 &  -20.11136290  & 38.27364540  \\
\vdots & \vdots &  \vdots &  \vdots & \vdots &  \vdots  & \vdots  \\
50 & 10.00000000 &  28.00000000 &  2.66666667 & -15.25561980 &  -20.11136290  & 38.27364540  \\	 	 
\hline
\end{tabular}
\end{center}
\end{sidewaystable}

After estimating the values of the parameters,  its  accuracy can be checked
as follows.  Compare the equilibrium point (for example $y^*_i=z^{(0)}_i$ if
we assume $u^{(0)}_i=0$ in Eq. (\ref{eqi_point})) assumed to be exhibited by
the modified system in the above procedure  with the equilibrium point
calculated from Eq. (\ref{mod_sys}) with the values of parameters determined
by the above method. The degree of closeness of these equilibrium points gives
a measure of accuracy of the estimated parameters. 

As an illustration to our method, let us consider the
Lorenz system
\bse
\label{lor_org}	
\bear
\dot{x}_1 &=& \sigma (x_2-x_1), 		\\
\dot{x}_2 &=& \rho x_1-x_2-x_1x_3,	 \\
\dot{x}_3 &=& x_1x_2-bx_3,
\eear
\ese
where $\sigma$, $\rho$ and $b$ are the unknown system (control) parameters. 
Then the modified Lorenz system can be constructed as
\bse
\label{lor_mod}	
\bear
\dot{y}_1 &=& \sigma (y_2-y_1) + \kappa_1,		 \\
\dot{y}_2 &=& \rho y_1-y_2-y_1y_3 + \kappa_2,		 \\
\dot{y}_3 &=& y_1y_2 -by_3 + \kappa_3,		
\eear
\ese
where $\kappa_1$, $\kappa_2$ and $\kappa_3$ are constants to be determined  
which make (\ref{lor_mod}) to exhibit equilibrium point solution while the
original system (with $\kappa_1=\kappa_2=\kappa_3=0$) exhibits chaotic
solution. Then the equation (\ref{dif_sys})  for the deviations,
$u_i=y_i-x_i$, $i=1,2,3$,  becomes
\bse
\label{lor_dif}	
\bear
\dot{u}_1 &=& \sigma (u_2-u_1) 	+ \kappa_1,	 \\
\dot{u}_2 &=& (\rho-x_3)u_1-u_2-(x_1+u_1)u_3 + \kappa_2,	  \\
\dot{u}_3 &=& (x_2+u_2)u_1+x_1u_2-bu_3 + \kappa_3. 
\eear
\ese
It may be noted that the presence of $x_1,\; x_2$ and $x_3$ in (\ref{lor_dif})
indicates that one has to solve Eqs.
(\ref{lor_org}) \& (\ref{lor_dif}) simultaneously.  \vskip .2cm
\begin{figure} 
\begin{center} 
\epsfig{figure=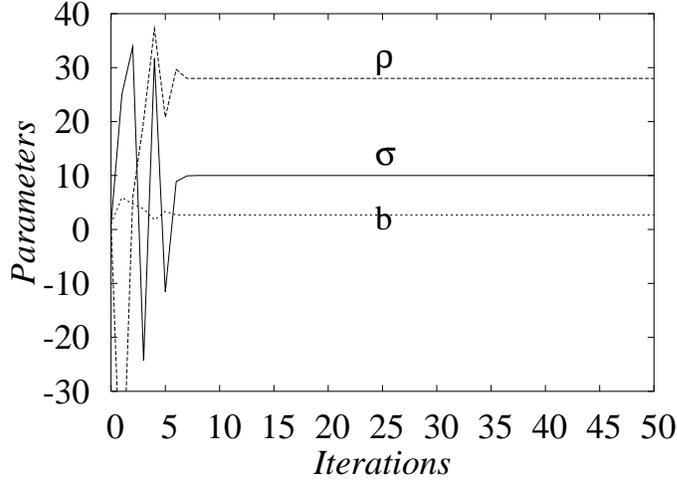,width=0.7\columnwidth} 
\end{center} 
\caption{Convergence of the parameters $\sigma$, $\rho$ and $b$ 
towards their exact values in the Lorenz system.}
\label{convergence}
\end{figure}

Let
$(z^0_1,z^0_2,z^0_3),(z^1_1,z^1_2,z^1_3),...,(z^{(m-1)}_1,z^{(m-1)}_2,z^{(m-1)}_3)$
be the  time series data obtained  from the Lorenz system at some arbitrary
time interval with the sampling rate $\Delta t$ for an unknown set of system
parameters.  After assuming $z^{(0)}_i$ as the equilibrium point for the
modified Lorenz system (that is by assigning $u^{(0)}_i=0$ in Eq.
(\ref{eqi_point}), so that $y^*_i=z^{(0)}_i$) and substituting three data
points $z^{(0)}_i$, $z^{(1)}_i$ and $z^{(2)}_i$ (one can in fact take any
three  successive data points but the first data point has to be used as
initial condition for $x_i$) in Eq. (\ref{eqi_point}), we get

\bear
&& z^{(1)}_1+u^{(1)}_1 = z^{(0)}_1 ,\;\;\;\;
z^{(1)}_2+u^{(1)}_2 = z^{(0)}_2 ,\;\;\;\;
z^{(1)}_3+u^{(1)}_3 = z^{(0)}_3 , \nonumber \\
&& z^{(2)}_1+u^{(2)}_1 = z^{(0)}_1 ,\;\;\;\;
z^{(2)}_2+u^{(2)}_2 = z^{(0)}_2 ,\;\;\;\;
z^{(2)}_3+u^{(2)}_3 = z^{(0)}_3. 
\label{lor_eqi}
\eear

Note that the time derivatives of $u_i$'s have specific  functional relations
with the unknown parameters ($\sigma$, $\rho$ and $b$) and controlling
constants ($\kappa_1$, $\kappa_2$ and $\kappa_3$) through Eq. (\ref{lor_dif}),
and hence  $u_i$'s are also  functions of the same parameters and controlling
constants.  Now we have six (implicit) functional relations for the six
unknowns namely $\sigma$, $\rho$, $b$, $\kappa_1$, $\kappa_2$ and $\kappa_3$
through Eq. (\ref{lor_eqi}), in terms of the equilibrium point (here
$z^{(0)}_i$) and the two sets of data points $z^{(1)}_i$ and  $z^{(2)}_i$,
$i=1,2,3$.  To solve for these unknowns, we have to obtain values of
$u^{(1)}_1$, $u^{(1)}_2$, $u^{(1)}_3$, $u^{(2)}_1$, $u^{(2)}_2$ and $u^{(2)}_3$
by solving Eqs. (\ref{lor_org})\&(\ref{lor_dif}) simultaneously so that  Eq.
(\ref{lor_eqi}) is satisfied. This can be done, for example, by the globally
convergent Newton's method \cite{nm} provided an initial guess is given to
all the unknown parameters.  In this example, while obtaining the value of
$u^{(1)}_i$ initial conditions should be  taken as $x^{(0)}_i=z^{(0)}_i$ and
$u^{(0)}_i=0$,  respectively. Further, the evaluation of $u^{(2)}_i$ needs to
reset $x^{(1)}_i=z^{(1)}_i$  in the numerical algorithm used above.  Similar
procedure has to be followed if any other set of successive data is used in
place of $z^{(0)}_i$, $z^{(1)}_i$ and $z^{(2)}_i$ in Eq. (\ref{lor_eqi}).  For
illustration purpose, we have used the numerically generated time series of the
Lorenz system for the system parameters $\sigma=10.0$, $\rho=28.0$ and $b=8/3$
and solved the Eq. (\ref{lor_eqi})  by  globally convergent Newton's method
with an initial guess  $1.0$ to all the unknowns $\kappa_1$, $\kappa_2$,
$\kappa_3$, $\sigma$, $\rho$ and $b$.   The convergence of the system
parameters and controlling constants ($\kappa$'s) is shown in the Table I,
which shows that the estimated values are indeed the exact values of the
parameters at which the time series data of the Lorenz system is generated.  We
also note that the convergence is very rapid, which is further confirmed in
Fig. (\ref{convergence}).  Note that if we take a different set of data points
from the time series as the equilibirium point, the corresponding controlling
contstants ($\kappa_i$'s) will  be different eventhough the system parameters
are unaltered.

\begin{figure} 
\begin{center} 
\epsfig{figure=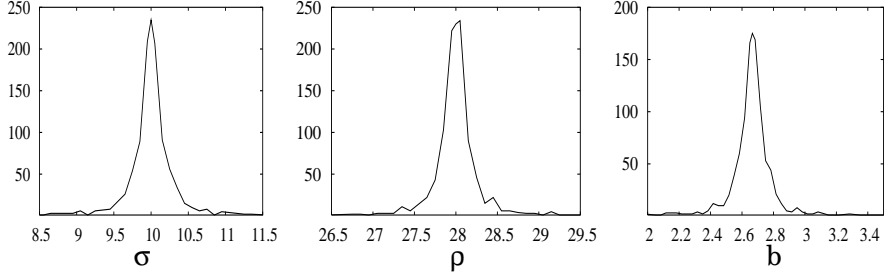,width=0.9\columnwidth} 
\end{center} 
\caption{The distribution of the values of the system parameters estimated
from the time series which contains random dynamical noise in the  
Lorenz system.}
\label{dis_dyn}
\end{figure}

\begin{figure} 
\begin{center} 
\epsfig{figure=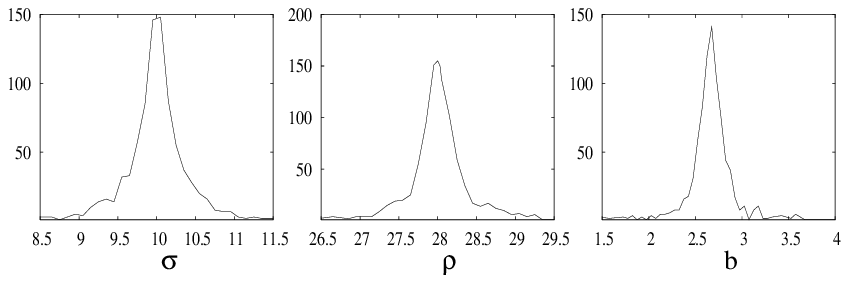,width=0.9\columnwidth} 
\end{center} 
\caption{The distribution of the values of the system parameters estimated
from the time series which contains random observational noise in 
the Lorenz system.}
\label{dis_obj}
\end{figure}

In order to show that our method is robust, a time series which contains
\emph{random dynamical noise} of strength $10^{-3}$ was additionaly introduced
in the above Lorenz model.   In this case, the distribution of the values of
the system parameters estimated using $1000$ data points is shown in fig.
(\ref{dis_dyn}).  Also in the case of \emph{random observational noise} of same
strength, the distribution of the values of the parameters is shown in fig.
(\ref{dis_obj}). The peaks about the true values of the parameters in the above
two figures indicate that our method works extremely well even in the presence
of dynamical noise or in the presence observational noise in the time series.

We have also carried out a similar analysis for the autonomous Chua's Circuit
~\cite{mlkm} in its dimensionless form,
\bse
\bear
\dot{x_1}&=&\alpha (x_2-x_1-h(x_1)), 	\\
\dot{x_2}&=&x_1-x_2+x_3, 			\\
\dot{x_3}&=&-\beta x_2,
\eear
\ese
where
\be
h(x)=bx+0.5(a-b)(|x+1|-|x-1|), \nonumber
\ee
$\alpha$ and  $\beta$ are parameters to be determined, and 
for the R\"ossler system,
\bse
\bear
	\dot{x_1} &=& -(x_2+x_3), 	\\	 
	\dot{x_2} &=& x_1+a x_2,  	\\
	\dot{x_3} &=& b+(x_1-c)x_3,
\eear
\ese
where $a$, $b$ and $c$ are the parameters to be determined.  In both the cases,
we are able to obtain the correct values of the parameters as in the case of the 
Lorenz system.
 
Next, we wish to point out that it is also possible to extend the analysis to 
predict the flow function of the system itself in principle, by assuming a 
polynomial form (here as an illustration) for the functions
$f_i(x_1,x_2,...x_N;{\bf p})$ in the right hand side of Eq. (\ref{org_sys}),
with unknown coefficients, and  solving sufficient number of Eqs.
(\ref{eqi_point}) to determine them. One can as well choose other types basis
functions and our procedure is equally applicable here also. To illustrate this
idea, let us consider the same time series data used in the above example of
Lorenz system and assume that the form of the underlying dynamical equations is
unknown. Let us assume a general quadratic form for the flow function
(dynamical equations) as 

\bse
\label{eqest_org}	
\bear
\dot{x}_1 &=& c_1x_1+c_2x_2+c_3x_3 	 
+c_4x_1x_2+c_5x_2x_3+c_6x_1x_3,			\\	 
\dot{x}_2 &=& c_7x_1+c_8x_2+c_9x_3	 
+c_{10}x_1x_2+c_{11}x_2x_3+c_{12}x_1x_3,	 	\\
\dot{x}_3 &=& c_{13}x_1+c_{14}x_2+c_{15}x_3	 
+c_{16}x_1x_2+c_{17}x_2x_3+c_{18}x_1x_3,
\eear
\ese
where $c_i$'s, $i=1,2,...18$, are parameters to be determined from the time series
data.  Again following the method suggested above, we write the form of the
modified system as 
\bse
\label{eqest_mod}	
\bear
\dot{y}_1 &=& c_1y_1+c_2y_2+c_3y_3	
+c_4y_1y_2+c_5y_2y_3+c_6y_1y_3+\kappa_1,			\\	 
\dot{y}_2 &=& c_7y_1+c_8y_2+c_9y_3	
+c_{10}y_1y_2+c_{11}y_2y_3+c_{12}y_1y_3+\kappa_2,	 \\
\dot{y}_3 &=& c_{13}y_1+c_{14}y_2+c_{15}y_3	
+c_{16}y_1y_2+c_{17}y_2y_3+c_{18}y_1y_3+\kappa_3,	
\eear
\ese
where $\kappa_1$, $\kappa_2$ and $\kappa_3$ are again the controlling constants
to be determined  so that  the above modified dynamical system exhibits a
stable equilibrium point.  For this system, Eq. (\ref{dif_sys}) becomes
\bse
\label{eqest_dif}	
\bear
\dot{u}_1 &=& \kappa_1+c_1u_1+c_2u_2+c_3u_3 	 
               +(c_4x_2+c_6x_3+c_4u_2)u_1		\nonumber 	\\
     	     &&+(c_4x_1+c_5x_3+c_5u_3)u_2		 
     	       +(c_5x_2+c_6x_1+c_6u_1)u_3,		 \\
\dot{u}_2 &=& \kappa_2+c_7u_1+c_8u_2+c_9u_3	 
               +(c_{10}x_2+c_{12}x_3+c_{10}u_2)u_1	\nonumber 	\\
     	     &&+(c_{10}x_1+c_{11}x_3+c_{11}u_3)u_2	 
    	       +(c_{11}x_2+c_{12}x_1+c_{12}u_1)u_3,	 \\
\dot{u}_3 &=& \kappa_3+c_{13}u_1+c_{14}u_2+c_{15}u_3	 
               +(c_{16}x_2+c_{18}x_3+c_{16}u_2)u_1	\nonumber 	\\
     	     &&+(c_{16}x_1+c_{17}x_3+c_{17}u_3)u_2		 
     	       +(c_{17}x_2+c_{18}x_1+c_{18}u_1)u_3.	 
\eear
\ese
In order to determine the values of the parameters and controlling constants
($c_i$, $i=1,2,...18$, $\kappa_1$, $\kappa_2$ and $ \kappa_3$), we have to
solve  $21$ algebraic equations which are actually constructed by making use of
eight successive time series data in Eq. (\ref{eqi_point}) with assumption that
the first one is the equilibrium point.  Then the required equations will be 
\bear
z^{(j)}_i+u^{(j)}_i &=& z^{(0)}_i, \;\;\;\;
i=1,2,3  \;\; \text{and} \;\; j=1,2,...,7. 
\label{eqest_eqi}
\eear

\begin{figure} 
\begin{center} 
\includegraphics[width=.9\linewidth]{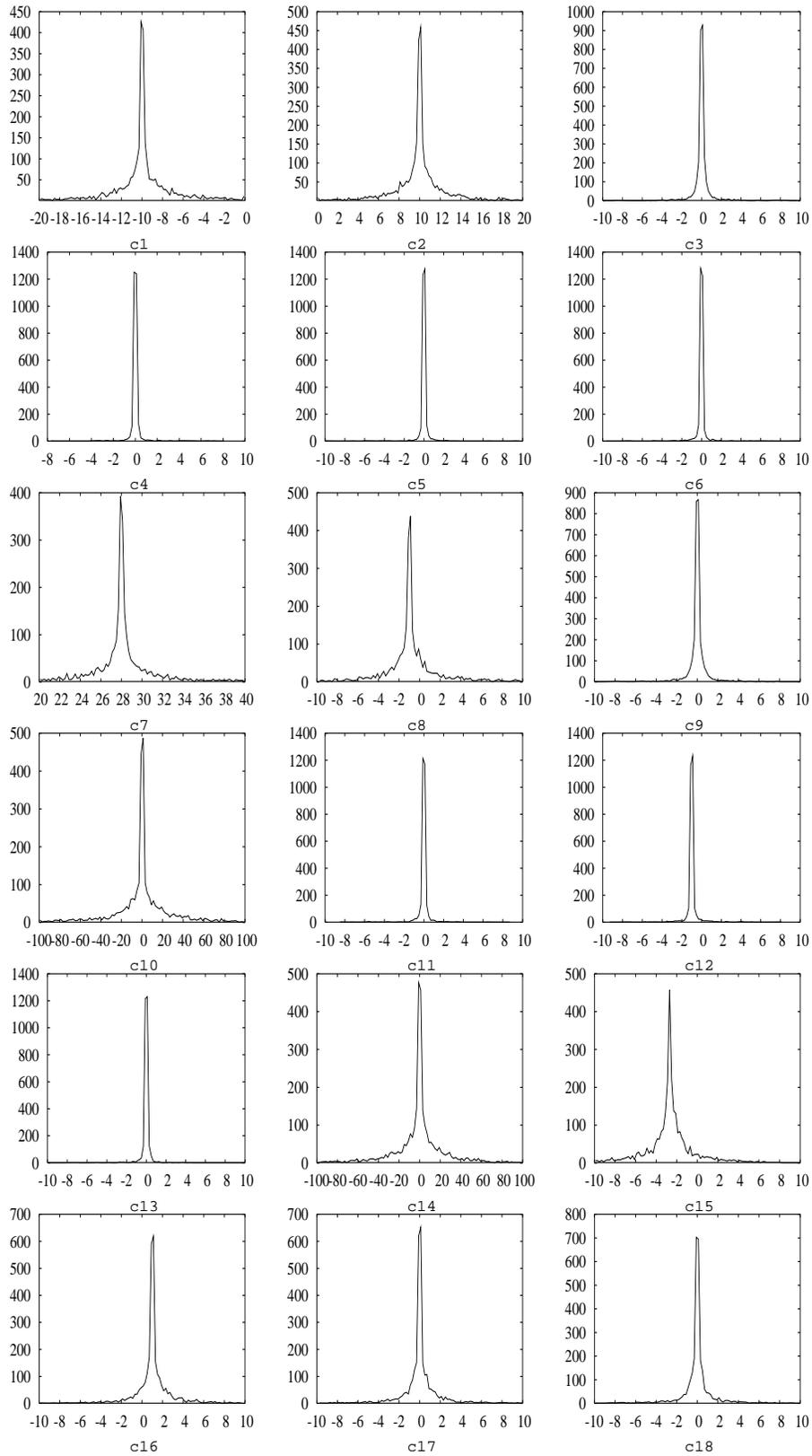}
\end{center} 
\caption{The distribution of the values of the system parameters of the flow
functions given by Eq. (\ref{eqest_org}), estimated
from the time series of Lorenz system}
\label{dis_eqest}
\end{figure}

Again we follow exactly the steps mentioned earlier for the Lorenz system and
obtain the values of the unknown parameters.  The distribution of the values of
the parameters ($c_i$'s, $i=1,2,...,18$) obtained by  solving Eq.
(\ref{eqest_eqi}) using $3000$ data of time series is shown is Fig.
(\ref{dis_eqest}). And the values are found to be distributed around
$\{c_i\}^{18}_1=\{-10$, $10$, $0$, $0$, $0$, $0$, $28$, $-1$, $0$,  $0$, $0$,
$-1$, $0$, $0$, $-2.67$, $1$, $0$, $0\}$.  Substituting these values in Eq.
(\ref{eqest_org}) we obtain the flow function  (evolution equation) of the form
\bse
\label{eqest_form}	
\bear
\dot{x}_1 &=& c_1x_1+c_2x_2,		\\
\dot{x}_2 &=& c_7x_1+c_8x_2+c_{12}x_1x_3,	 \\
\dot{x}_3 &=& c_{15}x_3+c_{16}x_1x_2,
\eear
\ese
with $\sigma=-c_1=c_2=10.0$, $\rho=c_7=28.0$, $b=-c_{15}=2.67$ and the
remaining constants $c_{16}=1.0$ and $c_8=c_{12}=-1$, which is nothing but the
original Lorenz system.

Finally, we would like to point out that the procedure outlined above also
gives a method to obtain the values of  the controlling constants ($\kappa_i$)
for a chaotic system to be controlled to a desired equilibrium point if the
form of evolution equation is known. 

To conclude, we have developed a very simple as well as  useful method for
estimating the unknown system parameters of the continuous dynamical systems of
any dimensions from the vector time series, if partial information is known,
namely the form of the dynamical equations .  It has also been extended to
obtain the form of the system equation itself at least in the case of
polynomial forms. Both the methods have been illustrated by means of vector
time series collected from the Lorenz system, while further confirmation has
been made with other systems including the Chua's Circuit and R\"ossler
systems. We have also checked that the method is robust againt dynamical and
observational noise and that the procedure exhibits a rapid convergence. 

\ack{This work has been supported by the National Board of Higher Mathematics,
Department of Atomic Energy, Government of India and the Department of Science
and Technology,  Government of India through research projects.
}

\end{document}